\documentclass{emulateapj}

\newcommand{\A}{{\rm \AA}}
\newcommand{\Teff}{T_{\rm eff}}
\newcommand{\BD}{BD$+44^\circ493$}

\shorttitle{The Ninth Magnitude Carbon-Enhanced Metal-Poor Star \object{\BD}}
\shortauthors{Ito et al.}

\begin{document}

\title{\object{\BD}: A Ninth Magnitude Messenger from the Early Universe;\\
Carbon Enhanced and Beryllium Poor}

\author{Hiroko Ito\altaffilmark{2,3}, Wako Aoki\altaffilmark{2,3}, Satoshi Honda\altaffilmark{4} and Timothy C. Beers\altaffilmark{5}}

\altaffiltext{1}{Based on data collected at the Subaru Telescope, which is operated by the National Astronomical Observatory of Japan.}
\altaffiltext{2}{Department of Astronomical Science, School of Physical Sciences, The Graduate University of Advanced Studies, Mitaka, Tokyo 181-8588, Japan; hiroko.ito@nao.ac.jp.}
\altaffiltext{3}{National Astronomical Observatory of Japan, Mitaka, Tokyo 181-8588, Japan; aoki.wako@nao.ac.jp.}
\altaffiltext{4}{Gunma Astronomical Observatory, Agatsuma, Gunma 377-0702, Japan; honda@astron.pref.gunma.jp.}
\altaffiltext{5}{Department of Physics \& Astrophysics, Center for the Study of Cosmic Evolution, and Joint Institute for Nuclear Astrophysics, Michigan State University, East Lansing, MI 48824-1116; beers@pa.msu.edu.}

\begin{abstract}
We present a 1D LTE chemical abundance analysis of the very bright ($V=9.1$) Carbon-Enhanced Metal-Poor (CEMP) star \object{\BD}, based on high-resolution, high signal-to-noise spectra obtained with Subaru/HDS. The star is shown to be a subgiant with an extremely low iron abundance ([Fe/H] $=-3.7$), while it is rich in C ([C/Fe] $=+1.3$) and O ([O/Fe] $=+1.6$). Although astronomers have been searching for extremely metal-poor stars for decades, this is the first star found with [Fe/H] $<-3.5$ and an apparent magnitude $V<12$. Based on its low abundances of neutron-capture elements (e.g., [Ba/Fe] $=-0.59$), \object{\BD} is classified as a "CEMP-no" star. Its abundance pattern implies that a first-generation faint supernova is the most likely origin of its carbon excess, while scenarios related to mass loss from rapidly-rotating massive stars or mass transfer from an AGB companion star are not favored. From a high-quality spectrum in the near-UV region, we set an very low upper limit on this star's beryllium abundance ($A({\rm Be}) = \log({\rm Be/H})+12 < -2.0$), which indicates that the decreasing trend of Be abundances with lower [Fe/H] still holds at [Fe/H] $<-3.5$. This is the first attempt to measure a Be abundance for a CEMP star, and demonstrates that high C and O abundances do not necessarily imply high Be abundances.
\end{abstract}


\section{Introduction}

Chemical abundance analyses of metal-poor stars based on high-resolution spectroscopy play an important role in our understanding of the chemical evolution of the early universe (e.g., \citealt{Beers05}) because their atmospheres preserve (in most cases) the chemical composition of the gas from which they formed. In particular, the nucleosynthesis signatures from the first generations of (likely massive) stars are believed to be recorded by lower-mass stars at the lowest observed metallicity. However, the number of extremely metal-poor stars having [Fe/H] $<-3.5$ known to date is very small ($\sim15$), and the chemical-enrichment processes that were in operation at such low metallicity are still unclear.

Surveys for metal-poor stars over the past few decades have shown that a large fraction of metal-poor stars exhibit enhancements of carbon. Such stars are often referred to as Carbon-Enhanced Metal-Poor (CEMP) stars. In particular, all three ultra metal-poor ([Fe/H] $<-4$) stars discovered so far exhibit strong C excesses. Several sources that could produce a significant amount of carbon in the early universe have been suggested to account for these observations. For example, mass transfer from an asymptotic giant branch (AGB) companion is proposed to explain the so-called ``CEMP-{\it s}'' stars, which have excesses of {\it s}-process elements as well as carbon. However, some CEMP stars, in particular for [Fe/H] $<-2.6$ \citep{Aoki07}, are not rich in {\it s}-process elements (``CEMP-no''). Among the handful of CEMP stars with [Fe/H] $<-3.5$ studied to date, none exhibits a clear excess of the {\it s}-process elements. One must therefore consider other scenarios where the C excess at extremely low metallicity may have been produced by previous generations of stars, e.g., mass loss from rapidly-rotating massive stars \citep{Meynet06} or the ejecta of so-called ``faint'' supernovae \citep{Umeda05}. In order to distinguish between these, and other possible scenarios, measured abundances of N and O, as well as for the heavy neutron-capture elements, are required.

In this Letter we present a 1D LTE abundance analysis of the extremely metal-poor ([Fe/H]$=-3.7$) CEMP star \object{\BD}. \cite{ATT94} reported the metallicity of this object to be [Fe/H]$=-2.71$, based on {\it uvby} photometry, while \cite{Carney03} pointed out that a synthetic spectrum with lower metallicity produced a better fit to the spectrum they used to measure its radial velocity. Our analysis reveals the presence of strong CH bands, which might lead to the overestimate of this star's iron abundance, and that the lower metallicity speculation by \cite{Carney03} is indeed correct. \object{\BD}, with apparent magnitude $V=9.1$, is quite remarkable in that it has the lowest iron abundance of any star with $V<12$ found to date, and is by far the brightest extremely low-metallicity star among all objects that have reported [Fe/H] based on high-resolution spectroscopy (according to the SAGA database; \citealt{Suda08}).

Thanks to its brightness, our high $S/N$ UV spectrum also allows inspection of the strength of the \ion{Be}{2} lines at 3130$\,\A$ for \object{\BD}, permitting measurement of a meaningful upper limit for beryllium at the lowest metallicity yet achieved. In the past decade, observations have revealed a clear linear correlation between stellar Be abundances and metallicity (e.g., \citealt{Boesgaard99}). This result was at first surprising, because standard cosmic-ray spallation processes, in which high-energy protons and $\alpha$-particles accelerated by supernovae impact interstellar CNO, predicted a quadratic correlation (``secondary'' process). The observations indicated the need for a ``primary'' process, in which accelerated CNO nuclei impinge on interstellar protons and $\alpha$-particles (e.g., \citealt{Yoshii97}). Such a primary process is expected to be especially important in the early universe, due to the lack of large amounts of interstellar CNO. The mechanism to accelerate CNO nuclei is still uncertain, but superbubbles (e.g., \citealt{Parizot99}) and type Ic supernovae (e.g., \citealt{Nakamura04}) have been suggested as possibilities. Our analysis of Be for \object{\BD} provides new insight for the origin of Be in the early universe, as well as a constraint on models that predict a low-metallicity plateau of Be abundances, including the Inhomogeneous Big Bang Nucleosynthesis (IBBN) scenario \citep{Orito97}.

\section{Observations}

High-resolution spectroscopy of \object{\BD} was carried out on 2008 October 5 with HDS \citep{Noguchi02} on the Subaru Telescope for three grating settings, covering 3100-9350$\,\A$ with a resolving power of $R\sim90,000$ (no on-chip binning). The exposure time for the UV observation is 120 minutes, while those for the other settings used are 15 minutes. In order to obtain a spectrum over 3900-3950$\,\A$, which fell in the CCD gap in the October run, an additional exposure of 5 minutes was made with the same instrument on 2008 November 15, with a resolving power of $R\sim60,000$ and $2\times2$ on-chip binning.

Data reduction was performed with the IRAF\footnote{IRAF is distributed by the National Optical Astronomy Observatories, which is operated by the Association of Universities for Research in Astronomy, Inc. under cooperative agreement with the National Science Foundation.} echelle package. The $S/N$ ratio per pixel achieved was $\sim100/1$ at $3100\,\A$, $\sim400/1$ at $4500\,\A$, and $\sim250/1$ at $5000\,\A$. The spectrum covering the range 3400-6800$\,\A$, where the $S/N$ is the highest and many atomic lines exist, is used for the abundance analysis for most elements. We use the UV spectrum ($<3400\A$) to measure the Be lines and the molecular features of NH and OH. Equivalent widths are measured by fitting Gaussian profiles to isolated atomic lines. For blended lines and molecular features, the spectrum synthesis technique is employed.

\section{Abundance Analysis}

We adopt the effective temperature $\Teff$ of $5510\,{\rm K}$ determined by \cite{Carney03}, who used the color-$\Teff$ relations derived by \cite{Alonso99} based on the infrared flux method. Although they adopted a higher metallicity of [Fe/H] $=-2.7$ in the calculation, the results hardly change even if a lower metallicity is assumed. The surface gravity is obtained from the LTE ionization equilibrium between \ion{Fe}{1} and \ion{Fe}{2}, and the microturbulent velocity is determined from 107 \ion{Fe}{1} lines by demanding that no trend of Fe abundances with equivalent widths is found. For this analysis, as well as for the following measurements of elemental abundances, a 1D LTE abundance analysis was performed using the grid of ATLAS9 NEWODF model atmospheres \citep{Kurucz93,Castelli03}. The results are $\log g=3.7$ and $v_t=1.3\,{\rm km ~s^{-1}}$; the derived metallicity is [Fe/H] $=-3.68$. The NLTE effect on Fe I is estimated to be 0.1-0.3 dex by previous studies (e.g., \citealt{Asplund2005}). NLTE effects on determinations of abundances and stellar parameters will be discussed separately (Ito et al. in preparation). For elements for which no line is detected, we estimate upper limits based on the 3$\sigma$ error of equivalent width measurement, employing the formula given in \cite{Norris01}. The derived abundances and uncertainties are listed in Table~\ref{tbl1}.

\begin{deluxetable}{rrrrrrrrrr}
\tablewidth{0pc}
\tablecaption{Abundance Results and Uncertainties\label{tbl1}}
\tablehead{
 & & & & & \colhead{random} & \colhead{total} \\
\colhead{X} & \colhead{Ion} & \colhead{lines\tablenotemark{a}} & \colhead{$\log \epsilon({\rm X})$} & \colhead{[X/Fe]\tablenotemark{b}} & \colhead{error} & \colhead{error} }
\startdata
Li & 1 & Syn & 1.04 & ... & 0.03 & 0.09 \\
Be & 2 & Syn & $<-2.0$ & ... & ... & ... \\
C & CH & Syn & 6.02 & $+1.31$ & 0.10 & 0.23 \\
N & NH & Syn & 4.42 & $+0.32$ & 0.25 & 0.32 \\
O & OH & Syn & 6.57 & $+1.59$ & 0.15 & 0.26 \\
Na & 1 & 2 & 2.76 & $+0.27$ & 0.09 & 0.13 \\
Mg & 1 & 7 & 4.37 & $+0.52$ & 0.04 & 0.10 \\
Al & 1 & 3961$\A$ & 2.12 & $-0.57$ & 0.13 & 0.16 \\
Si & 1 & 3905$\A$ & 4.24 & $+0.41$ & 0.13 & 0.19 \\
Ca & 1 & 9 & 2.90 & $+0.27$ & 0.02 & 0.06 \\
Sc & 2 & 6 & $-0.20$ & $+0.43$ & 0.04 & 0.12 \\
Ti & 1 & 5 & 1.54 & $+0.32$ & 0.02 & 0.11 \\
Ti & 2 & 13 & 1.53 & $+0.31$ & 0.01 & 0.11 \\
V & 1 & 4379$\A$ & $<-1.00$ & $<-0.42$ & ... & ... \\
Cr & 1 & 6 & 1.52 & $-0.44$ & 0.04 & 0.11 \\
Mn & 1 & 2 & 0.59 & $-1.22$ & 0.09 & 0.14 \\
Fe & 1 & 107 & 3.77 & $-3.68$ & 0.01 & 0.11 \\
Fe & 2 & 7 & 3.77 & $-3.68$ & 0.03 & 0.11 \\
Co & 1 & 6 & 1.72 & $+0.48$ & 0.02 & 0.10 \\
Ni & 1 & 5 & 2.59 & $+0.04$ & 0.03 & 0.11 \\
Zn & 1 & 4722$\A$ & $<1.29$ & $<0.37$ & ... & ... \\
Sr & 2 & 2 & $-1.02$ & $-0.26$ & 0.09 & 0.15 \\
Y & 2 & 3774$\A$ & $-1.81$ & $-0.34$ & 0.13 & 0.17 \\
Zr & 2 & 4149$\A$ & $<-1.31$ & $<-0.22$ & ... & ... \\
Ba & 2 & 2 & $-2.10$ & $-0.59$ & 0.09 & 0.15 \\
Eu & 2 & 3819$\A$ & $<-3.03$ & $<+0.13$ & ... & ... \\
Pb & 1 & 4057$\A$ & $<-0.27$ & $<+1.41$ & ... & ...
\enddata
\tablenotetext{a}{``Syn'' indicates the use of spectrum synthesis. In the case that the number of lines used is one, as well as for upper limits, the wavelength of the line is given.}
\tablenotetext{b}{Solar abundances were taken from \cite{Asplund05}. For iron [Fe/H] is given.}
\end{deluxetable}

\begin{figure}
\plotone{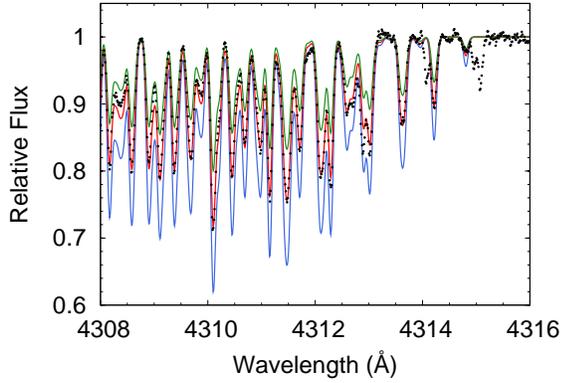}
\caption{The observed spectrum ({\it dots}) and synthetic spectra ({\it lines}) for the CH band at 4312$\,\A$. Assumed abundances are [C/Fe] $=+1.16$ ({\it blue line}), $+1.36$ ({\it red line}), and $+1.56$ ({\it green line}). \label{fig1}}
\end{figure}

We measure C abundances from the CH G-band at 4312$\,\A$ (Fig.~\ref{fig1}) as well as the feature at 4324$\,\A$. The CH line list of \cite{Aoki02} is used. The two C abundances differ by about 0.1 dex. We take the simple average to obtain our final C abundance, [C/Fe] $=+1.3$, which indicates that \object{\BD} is a carbon-enhanced object.

\begin{figure}
\plotone{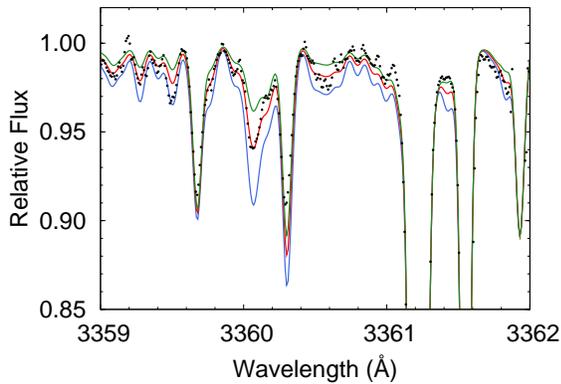}
 \caption{The same as Figure~\ref{fig1}, but for the NH band at 3360$\,\A$. Assumed abundances are [N/Fe] $=+0.12$ ({\it blue line}), $+0.32$ ({\it red line}), and $+0.52$ ({\it green line}). The N abundance is determined from the absorption feature at 3360.1$\,\A$. \label{fig2}}
\end{figure}

The N abundance is determined from the NH band feature at 3360$\,\A$. The synthetic spectra are calculated using the NH line list of \cite{Kurucz93} (Fig.~\ref{fig2}). \cite{Aoki06} found that the {\it gf}-values of the \cite{Kurucz93} NH line list needed to be corrected to reproduce the solar N abundance of \cite{Asplund05}. The same correction is applied to our analysis, with the result that N is not over-abundant in \object{\BD} ([N/Fe] $=+0.3$).

We measure O abundances via spectrum synthesis to match the OH features at 3123.95, 3127.69, 3128.29, 3139.17, and 3140.73 $\A$. All values agree with each other within 0.2 dex; we adopt their straight average as our final O abundance. The result is [O/Fe] $=+1.6$, indicating that \object{\BD} has a strong excess of O. However, the \ion{O}{1} triplet at 7775$\,\A$ is not detected in our spectrum, resulting in an upper limit from this feature to be [O/Fe] $<+1.41$, which is lower than the O abundance derived from the OH lines. Such a problem was also found for the carbon-enhanced hyper metal-poor star \object{HE~1327-2326} by \cite{Frebel06}, and a significant 3D effect on the OH features was suggested. It should be noted that the C/O ratio, which is important for the discussion of the origin of the carbon excess for this star, is expected to be comparatively robust, because both C and O abundances are determined from molecular features.

The Li abundance is derived from the \ion{Li}{1} doublet at 6708$\,\A$ using the spectrum synthesis approach. The derived abundance, $A$(Li)=1.04, agrees well with the values found in other metal-poor subgiants of similar effective temperature (see Figure 5 in \citealt{Frebel07}).

\begin{figure}
\plotone{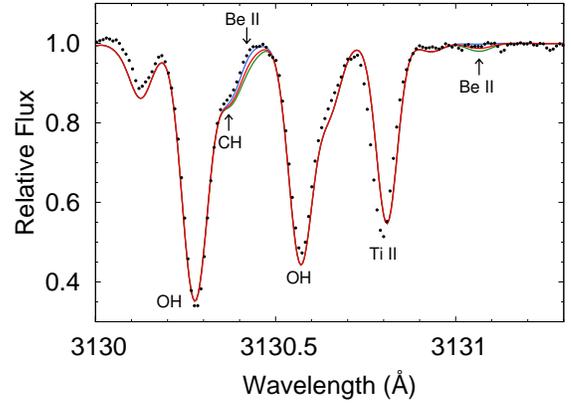}
\caption{The same as Figure~\ref{fig1}, but for OH and Be lines at 3131$\,\A$. Assumed abundances are $A({\rm Be})=-2.8$ ({\it blue line}), $-2.0$ ({\it red line}), and $-1.8$ ({\it green line}). In order to strictly estimate the Be abundance, two CH lines near the 3131$\,\A$ \ion{Be}{2} line are removed from the line list in the spectral synthesis. \label{fig3}}
\end{figure}

The \ion{Be}{2} resonance lines at 3130$\,\A$ are not detected. Using the line list based on \cite{Kurucz93} and \cite{Boesgaard99}, we synthesize spectra for different Be abundances and compare them with the observed spectrum (Fig.~\ref{fig3}). The upper limit estimated by this approach is $A({\rm Be}) < -2.0$. The upper limit estimated from the non-detection of 3131$\,\A$, employing the formula mentioned above, is $A({\rm Be}) < -2.18$. We adopted an upper limit of $A({\rm Be}) < -2.0$, which is the lowest Be abundance limit so far for metal-poor dwarfs or subgiants that have normal Li abundances.

The abundances of the $\alpha$-elements Mg, Si, Ca, and Ti are enhanced by $\sim$0.3-0.5 dex with respect to Fe, as found in most metal-poor stars of the halo.

No excess of neutron-capture elements is found. Note that the measurements (or upper limits) for Sr, Zr, and Y are all sub-solar. The Ba abundance ([Ba/Fe] $=-0.59$) is derived from the \ion{Ba}{2} resonance lines at 4554$\,\A$ and 4934$\,\A$, including the effect of hyperfine splitting using the line list of \cite{McWilliam1998} (assuming the isotope ratios of the {\it r}-process component of solar-system material). We only have an upper limit available for Eu, which is quite low as well ([Eu/Fe] $<+0.13$).

Random errors arising from the measurements of atomic lines are estimated to be $\sigma N^{-1/2}$, where $\sigma$ is the standard deviation of the individual derived abundances and $N$ is the number of lines used in the analysis. When the number of lines is smaller than five, we use the $\sigma$ of \ion{Fe}{1} instead. For molecular features, abundance variations from different features and continuum placement are taken into account to estimate random errors. In order to examine how stellar parameters affect abundances, we performed an abundance analysis after changing each stellar parameter by its uncertainty ($\Delta \Teff=100\,{\rm K} \,,\, \Delta \log g=0.3\,{\rm dex} \,,\, \Delta v_t =0.3\,{\rm km ~s^{-1}}$). The root-sum-square of all the error sources is adopted as the total error of $\log\epsilon({\rm X})$.

\section{Discussion}

From its over-abundance of C ([C/Fe] $>+1$) and low Ba abundance ([Ba/Fe] $<+0.5$), \object{\BD} is classified as a ``CEMP-no'' star, according to the nomenclature of \cite{Beers05}. The abundance pattern of \object{\BD} is very similar to that of \object{HE~1300+0157}, which is a CEMP-no subgiant with [Fe/H]$=-3.9$ \citep{Frebel07}, whose [C/Fe] and [O/Fe] are comparable with those of \object{\BD}. The brightness of \object{\BD} enables us to measure the abundances of N, Sr, Y, and Ba, which were not determined for \object{HE~1300+0157}. We can also set stronger upper limits for Zn, Pb, and so on. We now discuss the origin of the C excess in \object{\BD} and implications of the very low Be upper limit.

\begin{figure}
\plotone{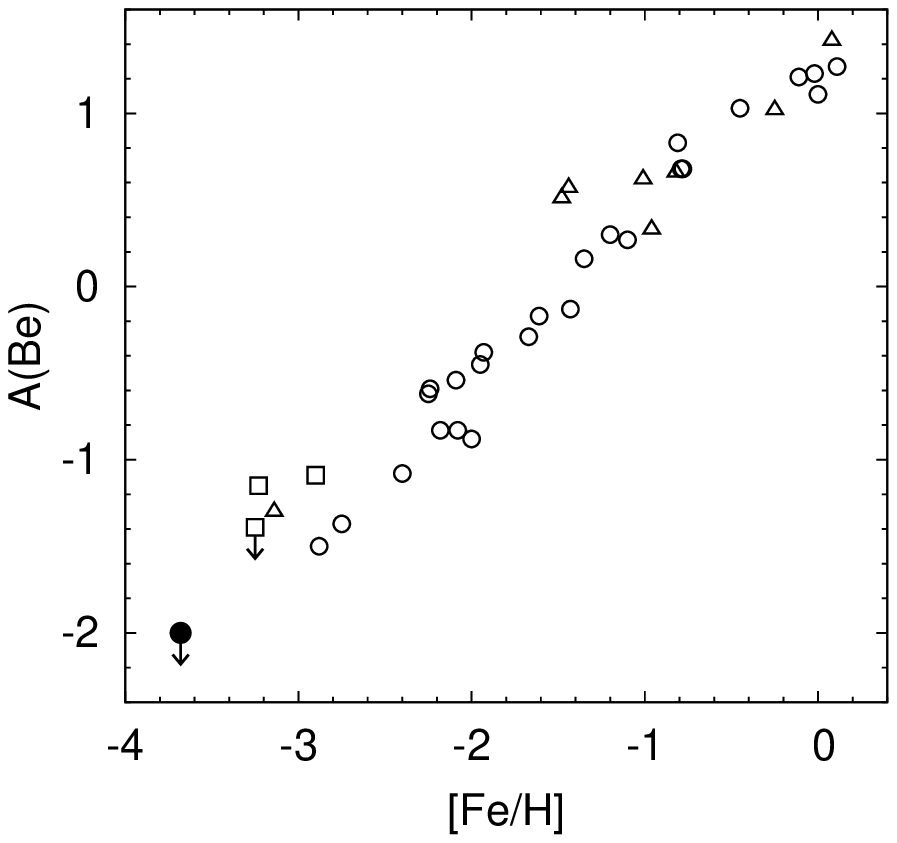}
\caption{$A$(Be) vs. [Fe/H]. Our result is the filled circle. Open circles and triangles are from \cite{Boesgaard99} and \cite{Boesgaard06}, respectively. Open squares indicate the results by \cite{Primas00a,Primas00b}. \label{fig4}}
\end{figure}

\subsection{The Origin of the Carbon Excess}

Since \object{\BD} is an unevolved subgiant, self-enrichment of carbon by dredge-up of the products of the helium core flash is clearly excluded.

For many CEMP stars, mass transfer from a companion AGB star is proposed as one of the causes of C enrichment, and such a model has had great success in explaining CEMP-{\it s} objects. However, the observations of \object{\BD} do not support this scenario. The first problem, which is common for CEMP-no objects, is that the neutron-capture elements that are expected to be enhanced by an AGB companion are not over-abundant in this star. \cite{Cohen06} suggested that the normal Ba abundances of CEMP-no stars can be explained by the high neutron-to-seed nuclei ratio in the {\it s}-process that occurred in the AGB companion, resulting in little Ba excess but a large Pb enhancement. Under this hypothesis, the Pb abundance predicted by them is $\log \epsilon({\rm Pb})=1.5$ at [Fe/H] $=-3.5$, which is much higher than our measured Pb upper limit for \object{\BD} ($\log \epsilon({\rm Pb})<-0.27$). \cite{Komiya07} discussed the possibility that a relatively high-mass companion AGB star ($M>3.5M_\odot$) produces little {\it s}-process elements due to inefficient radiative $^{13}{\rm C}$ burning. Since a high-mass AGB star converts C into N via hot-bottom burning, the low N abundance of \object{\BD} allows only a narrow range of mass. Another constraint is the low C/O ratio (C/O $<1$) found for \object{\BD}, which cannot be explained by the AGB nucleosynthesis scenario \citep{Nishimura09}. Moreover, radial velocity monitoring from 1984 to 1997 did not find any characteristic binarity signature \citep{Carney03}. The radial velocities measured by our data ($-150\,{\rm km ~s^{-1}}$ both in October and November 2008) agree with those in \cite{Carney03}. Our conclusion is that the binary mass-transfer scenario is excluded as the origin of the C excess of \object{\BD}.

Another scenario is that the C enhancement occurs prior to the formation of (at least some) CEMP stars due to the predicted mass loss from rapidly-rotating massive stars with very low metallicity. \cite{Meynet06} explored models of rapidly-rotating $60M_\odot$ stars with metallicities $Z=10^{-8}$ and $10^{-5}$, and found that stronger internal mixing increases the total metallicity at the stellar surface significantly, leading to a large mass loss. The ejecta are highly enriched in CNO elements, which are products of the $3\alpha$ reactions and the CNO cycle. In particular, the N excess is predicted to be quite large due to operation of the CNO cycle in the H-burning shell. However, the low observed N abundance of \object{\BD} is not explained by this scenario.

A remaining possibility is element production by so-called faint supernovae associated with the first generations of stars, which experience extensive matter mixing and fallback during their explosions \citep{Umeda03,Umeda05,Tominaga07b}. Such a process is realized in relativistic jet-induced supernovae with low energy deposition rates \citep{Tominaga07a}. The small amount of ejected $^{56}$Ni results in the faintness of the supernova, and high [C/Fe] and [O/Fe] ratios are predicted in the ejected material. This model can reproduce the observed abundance pattern of the CEMP star \object{HE~1300+0157} (Tominaga et ~al. in preparation), which is very similar to that of \object{\BD}. The abundances of elements that are newly determined for \object{\BD} (e.g., N) are also compatible with this model. The low [N/C] in \object{\BD} indicates that the mixing between the He convective shell and H-rich envelope during pre-supernova evolution is not significant \citep{Iwamoto05}. Thus, a faint supernova is the most promising candidate for the origin of the C excess in \object{\BD}.

\subsection{Implications of the Low Beryllium Abundance}

Figure~\ref{fig4} shows the Be abundance upper limit for \object{\BD} in the $A({\rm Be})$ vs. [Fe/H] plane, along with the results of previous studies. Our result is consistent with a linear trend between Be and Fe. Although Be is a fragile element, previous measurements for subgiants having $\Teff\sim5500\,{\rm K}$ showed no depletion. It is of significance that the Be abundance keeps decreasing and exhibits no plateau, at least to the level of $A({\rm Be})<-2.0$. Previous measurements of Be for stars with [Fe/H] $<-2.5$ permitted the possibility of a Be plateau around $A({\rm Be}) \sim -1.3$ \citep{Primas00b}, which would now be clearly called into question.

Models that have been invoked to account for a Be plateau include IBBN \citep{Orito97,Suzuki99}, interstellar accretion \citep{Yoshii95}, pre-Galactic cosmic rays generated by massive stars \citep{Kusakabe2008}, and a particular superbubble model \citep{Vangioni98}. Some of them predict a plateau as high as $A({\rm Be}) = -1.5 \sim -2.0$. The Be upper limit of \object{\BD} provides a meaningful constraint on these models, as it appears that any Be plateau, if it exists at all, must occur at $A({\rm Be})<-2.0$.

Our analysis is the first attempt to measure a Be abundance for a CEMP star. Since Be is produced via the spallation of CNO nuclei, their abundances, especially O abundances, have been expected to correlate with Be abundances. However, our low Be upper limit shows that the high C and O abundances in \object{\BD} are irrelevant to its Be abundance. Moreover, the origin of the C excess in CEMP-no stars like \object{\BD}, which we have argued is most likely due to faint supernovae, is unlikely to be a significant source of high-energy CNO nuclei that participate in the primary process associated with Be production.

\acknowledgments
W.~A. is supported by a Grant-in-Aid for Science Research from JSPS (grant 18104003). T.~C.~B. acknowledges partial funding of this work from grants PHY
02-16783 and PHY 08-22648: Physics Frontier Center/Joint Institute for Nuclear
Astrophysics (JINA), awarded by the U.S. National Science Foundation.

\end{document}